\documentclass[12pt]{article}

\usepackage{amsmath}
\usepackage{mathtools}
\usepackage{dashbox}
\usepackage{a4wide}
\usepackage{epsfig}
\usepackage{theorem}
\usepackage{amssymb}
\usepackage{bbm}
\usepackage[T1]{fontenc}
\usepackage[latin1]{inputenc}
\usepackage{cancel}
\usepackage{blkarray}
\usepackage{multirow}
\usepackage{tikz}
\newcommand{\bi}{\bigskip}
\newcommand{\no}{\noindent}
\newcommand{\bea}{\begin{eqnarray}}
\newcommand{\eea}{\end{eqnarray}}

\newcommand{\be}{\begin{equation}}
\newcommand{\ee}{\end{equation}}
\newcommand{\hk}{\hspace{0.1cm}}

\newcommand{\lk}{\left(}\usepackage{multirow}

\usepackage{amsmath}
\usepackage{a4wide}
\usepackage{epsfig}
\usepackage{theorem}
\usepackage{amssymb}
\usepackage{bbm}
\usepackage[T1]{fontenc}
\usepackage[latin1]{inputenc}

\begin{document}

\title{On Dirichlet's Derivation of the Ellipsoid Potential}

\author{W.~Dittrich\\
Institut f\"ur Theoretische Physik\\
Universit\"at T\"ubingen\\
Auf der Morgenstelle 14\\
D-72076 T\"ubingen\\
Germany\\
electronic address: qed.dittrich@uni-tuebingen.de
}
\date{\today}

\maketitle
\bi

\no

\begin{abstract}
Newton's potential of a massive homogeneous ellipsoid is derived via Dirichlet's discontinuous factor. 
At first we review part of Dirichlet's work in an English translation of the original German,  and then continue with an 
extension of his method into the complex plane. With this trick it becomes possible to first calculate the potential 
and thereafter the force components exerted on a test mass by the ellipsoid. This is remarkable in so far as all other
famous researchers prior to Dirichlet merely calculated the attraction components. Unfortunately, Dirichlet's derivation 
is to a large extent mathematically unacceptable which, however, can be corrected by treating the problem in the complex plane. 
\end{abstract}

\section{Introduction: The Homogeneous Ellipsoid}

The calculation of the attraction and the potential of the homogeneous ellipsoid
is one of the most-discussed problems in mathematical physics. Only at the beginning 
of the 19th century was a satisfactory solution found. It was at this time and the time
thereafter that the work of Laplace (1782), Ivory (1809), Gau\ss{} (1813), Chasles (1838) 
and Dirichlet (1839) took place \cite{1}.

This article is not so much about C.F. Gauss and his seminal contribution to the 
calculation of the attractive force components of a homogeneous massive ellipsoid 
in inner and outer space, but a homage to his successor in G\"ottingen, 
P.G. Lejeune-Dirichlet, who, for the first time, tried to compute the \underline{potential} 
prior to the \underline{force components}
that an ellipsoid exerts in inner and outer space. 

Gau\ss{}' contribution is even more important for another reason. In his article of 1813 
on the ellipsoid, he uses for the first time in history the divergence theorem which 
carries his name. The world of mathematics and physics would be unthinkable without
this integral theorem. True, the attraction components of the ellipsoid are calculated, 
but \underline{not} the potential, a concept that Gauss introduced later in 1840. 
However, from a 
handwritten remark which is re-printed in Vol. V, pp. 285-286 of Gau\ss{}' works, he 
lets us know that one can compute the potential as well using a method similar
to that employed in his ellipsoid paper.

The emphasis in the current article is on Dirichlet's so-called discontinuous factor 
which he uses to handle multiple integrals with function-like variables at the lower 
and upper limit of the integral. One should consider this idea as a ``predecessor''
to $\delta$-like functions (distributions) which were introduced by P.M.A. Dirac and are at 
the center of J. Schwinger's and others' treatment of any type of field theory with 
Green's functions in modern physics. It is precisely the $\delta$
function which can be represented by a limiting process of ``reasonable functions'' 
that can be used under an integral to pick out special values and forget about the
complicated upper and lower limits of the integral - exactly what Dirichlet had 
in mind when he discovered how easy it can be when computing multiple integrals as
in the case of the potential of a homogeneous massive ellipsoid. This is reason
enough to take a closer look at Dirichlet's trick to simplify complicated integrals 
introduced in 1839.

\section{On a New Method for Calculating Multiple Integrals
by
P.G.~Lejeune-Dirichlet}\label{section2}

Dirichlet's article, ``\"Uber eine neue Methode zur Bestimmung vielfacher Integrale'', 
was published  in extenso in Treatises of the Berlin Academy from the Year 1839, 
Berlin 1841 [pp. 61-79 in the Mathematical Papers]. 

The following represents an English translation of the introductory parts of Dirichlet's 
contribution, originally published in German.

It is well known that the calculation of a multiple integral or its reduction to a lower 
order is generally one of the more difficult problems encountered when the limits of
integration for the individual variables are not constant but are mutually dependent, 
so that the domain of integration is expressed by one or more inequalities containing 
more than one variable. When dealing with various physical problems which lead back 
to the calculation of a class of multiple integrals of an undetermined order, 
the author came across the method that is the subject of this article and which 
not only yields the value of the integral on which the present investigation relies, 
but also on the many other different kinds of integrals that it can be applied to. 
Nevertheless this method is so simple that one wonders why it hasn't been applied to 
similar studies well before now. The principle behind this approach to multiple integrals 
which are to be taken between variable limits is based on the well-known property of 
certain integrals which represent discontinuous functions of those constants that are 
contained in the integrals and are dependent in different intervals in various ways. 
For example, we know that the simple expression
\be
\label{2.1}
(\frac{2}{\pi}) 
\int\limits^\infty_0 \cos (g \varphi) \frac{\sin \varphi}{\varphi} d \varphi
\ee
is equal to unity as long as $g$ lies between $-1$ and $+1$, 
but disappears if $g$ lies outside this interval. If one has a three-fold integral - 
we are not considering one of higher order because with three variables the procedure 
takes on a geometric dimension, which allows us to describe the process - 
which is to extend over a defined space, e.g., over an ellipsoid surface, so that 
one can say that if 
$\alpha, \beta\ \gamma$ describe the semi-axes of this surface in which direction the 
coordinate axes coincide, the expression 
\be
\frac{x^2}{\alpha^2} + \frac{y^2}{\beta^2} + \frac{z^2}{\gamma^2} \nonumber
\ee
lies below or above unity, depending on whether the point $(x,y,z)$
lies  within or outside the specified space, to see immediately that the integral 
\be
\label{2.2}
(\frac{2}{\pi}) \int d \varphi \frac{\sin \varphi}{\varphi} \cos 
\left[ \left(\frac{x^2}{\alpha^2} + \frac{y^2}{\beta^2} + \frac{z^2}{\gamma^2} \right) \varphi
\right]    
\ee
has the value of unity inside of the ellipsoid, but disappears outside of it. 
So if one multiplies the given differential expression $Pdxdxdz$, where $P$
indicates some function of $x, y, z$, 
under the above integral in question, one no longer needs to take the original limits 
into account when integrating, i.e., one can perform the integrations with 
respect the variables $x, y, z$ between the constant 
limits $- \infty$ and $\infty$ in which, due to
the added discontinuous factors, the elements on which the integration should not
be extended drop out by themselves. This method can be described in two words 
in such a way that an integral extending in all directions over a limited mass 
distribution can be immediately transformed into one that stretches over an 
infinite space and in most cases will be much easier to handle because one permits 
the density outside of a given volume to become equal to zero; this can easily be done 
with a discontinuous factor. It is surprising to what extent the most difficult 
integrations can be easily performed with these transformations, which initially seem 
to hardly promise success, and how these problems, which often demand intricate and 
time-consuming calculations, can be solved without difficulty, simply with the help 
of a few well-known integrals.

In this paper we can only give a short description of a few of the applications of this
method. One example is the attraction of the ellipsoid, a problem that mathematicians 
have studied more than any other involving integral calculus.

Normally one reduces the case of a point external to the ellipsoid to an internal one, 
which is easier to calculate, or, if both are to be solved independently of each other,
then quite different means are used.

Using the above-described method, both cases can be treated in a similar manner and 
independently. First one must distinguish between the two in order to express the
result in a final and simple form. Furthermore, the procedure should not be limited
to the requirement that the attraction be inversely proportional to the distance 
squared, but rather remains applicable for any other integer or fractional 
power of the distance. Nor need the density of the attracting mass be assumed constant
but can be expressed by any rational, integer function of the coordinates $x, y, z$. 
For simplicity's sake, however, the density will be assumed to be constant and equal 
to unity.

Let $\alpha, \beta, \gamma$
be the semi-axes of the ellipsoids; $a, b, c$
the coordinates of the attracted point; and $x, y, z$
of any point of the attracting mass. Furthermore, let
\be
\rho^2 = (x-a)^2 + (y-b)^2 + (z-c)^2 \nonumber
             \ee
and $\frac{1}{\rho^p}$
be the law of attraction (where $p$
is assumed to lie between $2$ and $3$; 
beyond these limits the procedure requires a few minor modifications), then the force
component $A$ of the attraction parallel to the $x$ axis  
(and considered to be positive from the side where $x$'s  decrease), 
is obtained by taking the derivative  with respect to a of the integral covering the
entire ellipsoid:
\be
   \label{2.3}
   \frac{- 1}{(p-1)} \int \frac{dxdydz}{\rho^{p-1}} \,    .                                    
\ee
According to the above, the integral is transformed into
          \be
          \label{2.4}
\frac{- 2}{\pi (p-1)} \int\limits^\infty_0 d \varphi \frac{\sin \varphi}{\varphi} \int \cos
\left[ \lk \frac{x^2}{\alpha^2} + \frac{y^2}{\beta^2} + \frac{z^2}{\gamma^2} \right) \varphi 
\right] \frac{dxdydz}{\rho^{p-1}}                                                  
             \ee
where the integrations for $x, y, z$
can be extended from $- \infty$ to $\infty$. 
The calculation is much easier if, instead of this integral, we observe  the following 
one, whose real part coincides with that which we are looking for:
\be
\label{2.5}
\frac{- 2}{\pi (p-1)} \int\limits^\infty_0 d \varphi \frac{\sin \varphi}{\varphi} 
\int \exp \left[ i \varphi \lk \frac{x^2}{\alpha^2} + \frac{y^2}{\beta^2} + \frac{z^2}{\gamma^2}
\right) \right] \frac{dxdydz}{\rho^{p-1}} \, .                                                   
                    \ee
Integrating for $x, y, z$
cannot be performed in this form, but can easily be done if one expresses the factor 
$\frac{1}{\rho^{p-1}}$ with the help of a certain integral in such a way that the 
coordinates $y, y, z$, as in the other factor,   appear only in the exponents.

So much for the introduction to Dirichlet's original paper. Later in the text we will 
return to the entire calculation, then treating it in the complex plane. 
Before continuing, it would be appropriate to derive formula (\ref{2.1}) 
because this equation lies at the heart of Dirichlet's computational trick.

\section{Dirichlet's Discontinuous Integral -``Light''}

Let us start with the function $h(t)$, defined as
                                     \be
                                     \label{3.1}
                          h(t) =  \int\limits^\infty_0 e^{-xt} \frac{\sin(gx)}{x} dx \, , \quad \quad    t > 0, g = const.		                
                                           \ee          
                                           Differentiating with respect to the parameter $t$,
                                                     \be
\frac{dh}{dt} = - \int^\infty_0 e^{-xt} \sin(gx)dx\, , \nonumber
                                                                \ee                        
and integrating by parts twice, gives
\be
   \frac{dh}{dt} = - \frac{g}{(g^2+t^2) } \nonumber
\ee
which, when integrated, yields
\be
                              h(t) = A - \tan^{-1} ( \frac{t}{g}) \nonumber
\ee
with $A$ a constant of integration. From the integral definition of $h(t)$ (\ref{3.1}) 
we observe that $h(\infty) = 0$. Thus, $0 = C-\tan^{-1} (\pm \infty)$, 
where we use the + sign if $g > 0$ and the - sign if $g < 0$. Hence, $A = \pm \frac{\pi}{2}$ and we have
\be
\label{3.2}
                               h(t) = \pm \frac{\pi}{2} - \tan^{-1} (\frac{t}{g}) \, .                                
\ee
At this point we set $t = 0 $ and with it, $\tan^{-1} (\frac{t}{g}) = 0$, so that we obtain Dirichlet's discontinuous integral
\be
\label{3.3}                                                      
 \int\limits^\infty_0 \frac{\sin(gx)}{x} dx = \left\{ \begin{array}{clc} 
                                          \frac{\pi}{2} & \mbox{if} & g > 0    \\
                                              0 & \mbox{if} &  g = 0  \\
                                               - \frac{\pi}{2} & \mbox{if}  & g < 0
                                              \end{array}
                                  \right.
                          \ee                                             

 \begin{figure}
 \centerline{
 \includegraphics[width=.6\textwidth]{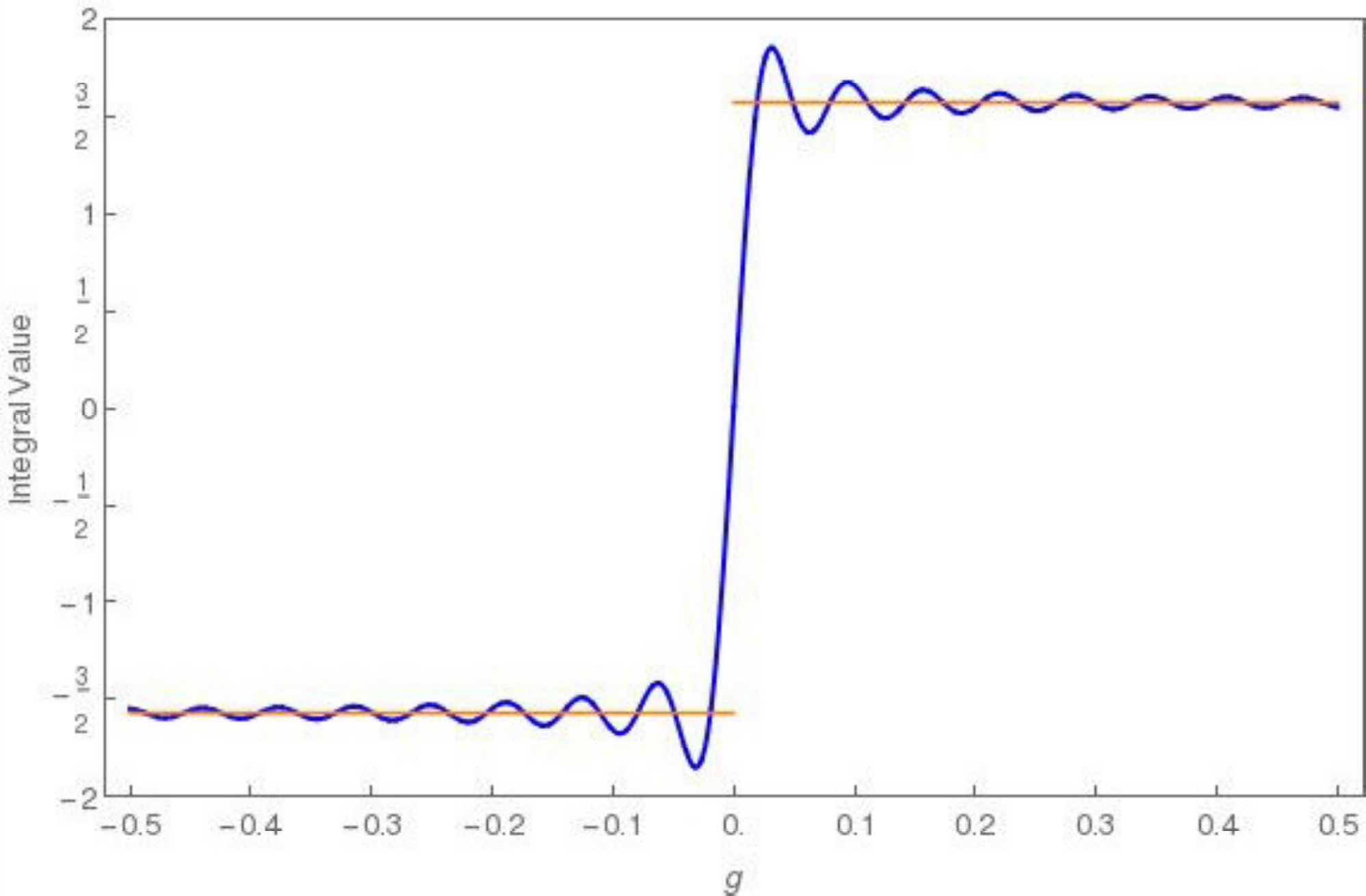}}
   \caption{Dirichlet's discontinuous integral}\label{Fig.1}
\end{figure}

The plot in Fig. \ref{Fig.1} shows Dirichlet's discontinuous integral with the sudden jump as g goes from negative to positive values.
By the way, Euler derived the special case $g =1$ at the end of his life in 1783:
\be
\label{3.4}
           \frac{2}{\pi} \int\limits^\infty_0 dx \frac{\sin x}{x} = 1 \, .
                                               \ee
This famous formula is sufficient to derive Dirichlet's expression of (\ref{2.1}). Here are two ways to prove this statement:
\begin{itemize}
 \item[(a)]
 Let us put $x = tu$ in (\ref{3.4}) where $t$ is a positive number. This gives
                                     \be
                                     \label{3.5}
\frac{2}{\pi} \int\limits^\infty_0 du \frac{\sin(tu)}{u}  = 1 \, .
                                               \ee
Now let $h>g$ be two positive numbers. Then $h+g$ and $h-g$
are also positive. If we substitute these two numbers for $t$ into the former integral, we obtain $(u=\varphi)$
                      \be
              \frac{2}{\pi} \int\limits^\infty_0 \frac{d \varphi}{\varphi} \sin(h+g)\varphi = 1 \, , \quad \quad
              \frac{2}{\pi} \int\limits^\infty_0 \frac{d \varphi}{\varphi} \sin(h-g)\varphi = 1 \, .\nonumber
                           \ee
Adding and subtracting these two integrals gives
\be
\frac{2}{\pi} \int\limits^\infty_0 \frac{d \varphi}{\varphi} \sin(h\varphi) \cos(g \varphi) = 1 \, , \quad \quad
\frac{2}{\pi} \int\limits^\infty_0 \frac{d \varphi}{\varphi} \sin(g \varphi) \cos(h \varphi) = 0 \, , \nonumber
              \ee
or
\be
\frac{2}{\pi} \int\limits^\infty_0 \frac{d \varphi}{\varphi} \sin(h \varphi) \cos(g\varphi) = 1 \, \mbox{or} \,  0 \, , \nonumber
                                        \ee
 depending on $h>g$ or $h<g$. Finally we put $h=1$ and so obtain                              
 \be
 \label{3.6}                                            
 \frac{2}{\pi} \int\limits^\infty_0 d\varphi \frac{(\sin \varphi)}{\varphi} \cos(g \varphi) =   
 \left\{ \begin{array}{clc}
          1 & \mbox{for} & g<1\\
           0 & \mbox{for} & g>1 \\
           \frac{1}{2} & \mbox{for} &  g = 1 
         \end{array}
\right. \, .                                     	      
\ee
Since the value of the integral in unchanged when we replace $+g$ by 
$-g$, we can also write                           
\be
\label{3.7}              
\frac{2}{\pi} \int\limits^\infty_0 d \varphi \frac{(\sin \varphi)}{\varphi} \cos (g \varphi) = 
 \left\{ \begin{array}{cll}
          1 & \mbox{for} & -1<g<1 \\
          0 & \mbox{for} & g<-1 \,  g>1 \\
          \frac{1}{2} & \mbox{for} & g = \pm 1
         \end{array}
\right. \, .                                     	      
\ee
Equation (\ref{3.6}) is precisely Dirichlet's discontinuity factor which is everywhere inside of an ellipsoid defined by
\be
\label{3.8}
                     g(x,y,z) = \frac{x^2}{\alpha^2} + \frac{y^2}{\beta^2} + \frac{z^2}{\gamma^2} < 1 \, ,     	               
                     \ee
equal to $1$ and equals zero for every point $(x,y,z)$ outside, i.e., 
\be
                     g(x,y,z) = \frac{x^2}{\alpha^2} + \frac{y^2}{\beta^2} + \frac{z^2}{\gamma^2} > 1.  \nonumber  
\ee
\item[(b)] Here is another derivation of Dirichlet's discontinuity factor. It starts with Fourier's integral formula
\begin{align}
 f(g) & = \frac{1}{\pi} \int\limits^\infty_0 d \varphi \int\limits^{+ \infty}_{- \infty} dtf(t) \cos \varphi (t-g) \nonumber\\
      &  = \frac{1}{\pi} \int\limits^\infty_0 d \varphi \cos (\varphi g) \int\limits^{+ \infty}_{- \infty} dtf(t) \cos (\varphi t) \nonumber\\    
      &   +\frac{1}{\pi} \int\limits^\infty_0 d \varphi  \sin (\varphi g) \int\limits^{+ \infty}_{- \infty} dtf(t) \sin (\varphi t) \, .
                                      \nonumber                                    \end{align}  
If $f(g)$ is an even function, we have
\be
\label{3.9}
f(g) = \frac{2}{\pi}  \int\limits^\infty_0 d \varphi \cos (\varphi g) \int\limits^\infty_0 dt f(t) \cos (\varphi t)  
\ee
and for odd function $f(g)$ we obtain
\be
   f(g) =  \frac{2}{\pi}  \int\limits^\infty_0 d \varphi \sin (\varphi g) \int\limits^\infty_0 dt f(t) \sin (\varphi t) \, . \nonumber                    
                                            \ee                 
Now, according to (\ref{3.7}),  Dirichlet's discontinuity function, $f(g)$ is even and
\begin{align}
\begin{array}{lllll}
f(g) & = & 1 & \mbox{for} &  g<1 \, , \nonumber\\
f(g) & = & \frac{1}{2} & \mbox{for} &  g = 1\, , \nonumber\\
f(g) & = & 0 & \mbox{for} &  g>1. \nonumber
\end{array}
\end{align}
Therefore we obtain from (\ref{3.9})
\be
      f(g) = \frac{2}{\pi}  \int\limits^\infty_0 d \varphi \cos (\varphi g) \int\limits^1_0 dtf(t) \cos (\varphi t) = \frac{2}{\pi} 
      \int\limits^\infty_0 d \varphi \cos (\varphi g)(\frac{\sin \varphi}{\varphi}) \, , \nonumber           
                      \ee
                      which is the desired result. 
                      \end{itemize}

\section{Dirichlet's Discontinuity Factor in the Complex Plane}

We could continue with section \ref{section2} and follow for the rest of the paper Dirichlet's original work where he generalizes Newton's law 
to
$r^{-p}$, in which $p$ is not necessarily an integer number.  In the course of the calculation, the exponent must be subjected to a 
two-fold limiting condition, so that for finite $p$ the interval $2<p<3$ is left over and the case of interest to us, $p = 2$
even, would not be permitted. This was shown in detail in Dirichlet's  lectures, published by G. ARENDT [3], 
where a further extension of the range of validity to $1<p<3$
is claimed, whose grounds are however not satisfactory. This is probably the ``unimportant modification''
that Dirichlet ([4] p. 404) mentions without further explanation.  The limitation to the attraction components also 
follows due to convergence difficulties when integrating; the potential itself is in fact not derived correctly by DIRCHLET; 
rather, a formula without proof or limits of validity for $p$ ([4] p. 408) is given. This ambiguity is probably also 
the reason that these elegant methods have not found entry into the textbook literature 
(including TISSERAND, M\'ec. c\'el. volume II). Wherever this was 
attempted ([5]), the derivation of the potential was dispensed with due to the above-mentioned difficulties, 
and only the force components were determined. This failing can be remedied if one assumes  
a complex formulation of the discontinuous factor rather than Dirichlet's version using e.g., the real Fourier integral. 
This is, by the way, also desirable, since DIRICHLET has to change over to complex integrals in the course of the calculation.
We define with real g>o as discontinuous factor  
\be
 \label{4.1}
		\frac{1}{\pi} \int\limits_C d \varphi e^{ig \varphi} \frac{(\sin \varphi)}{\varphi} = \left\{  1, g<1 \atop    0, g>1  \right.
		\ee
 \begin{figure}[h]
 \centerline{
 \includegraphics[width=.3\textwidth]{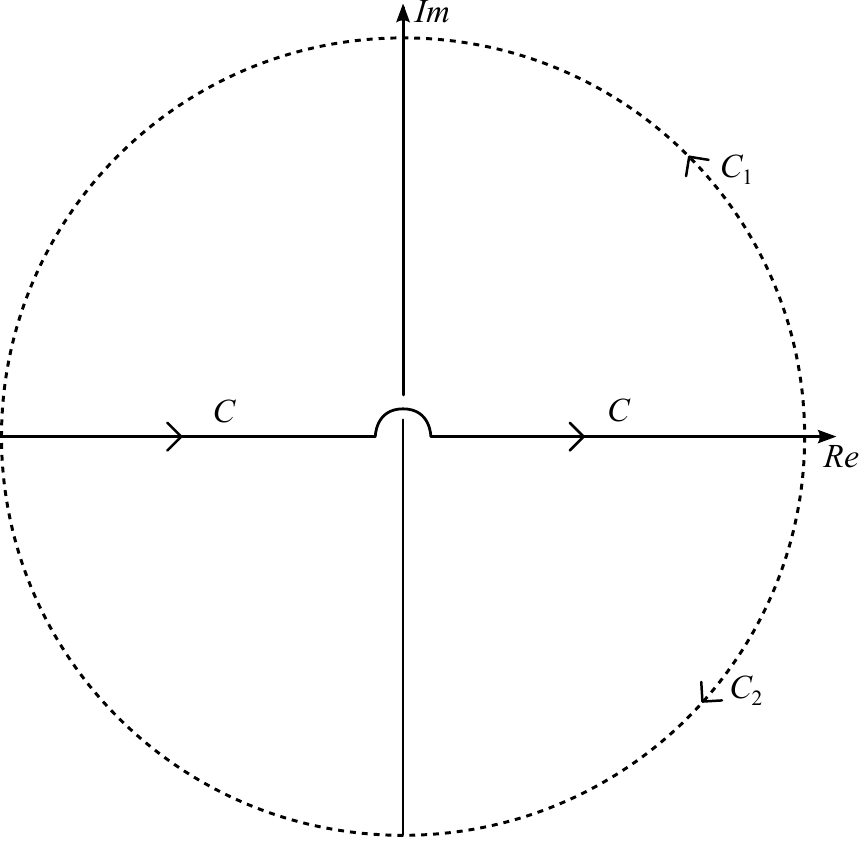}}
   \caption{Integration path in the complex $\varphi$ plane }\label{Fig.2}
\end{figure}
and take the whole real axis from $- \infty$ to $+ \infty$ as the integration path $C$, 
bypassing the zero point of the complex $\varphi$ plane by a small half-circle in the upper half-plane (see Fig. \ref{Fig.2}). 
Here is a proof of formula (\ref{4.1}).
\bi

\no
The path $C$ is given by \includegraphics[width=.1\textwidth]{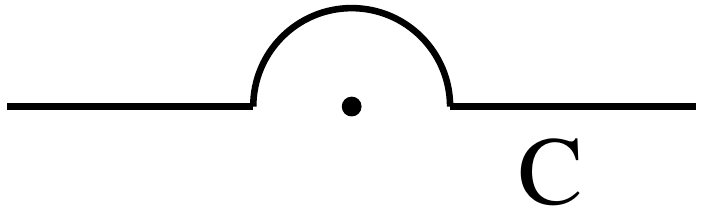} 
and $\sin \varphi = \frac{1}{(2i)} (e^{i \varphi} - e^{-i \varphi})$.
      \bi                                          

      \no
So we have to study the integral
\be
\label{4.2}
\frac{1}{\pi} \int\limits_C
d \varphi e^{ig \varphi}  \frac{\sin \varphi}{\varphi} = \frac{1}{\pi} \int\limits_C \frac{d \varphi}{\varphi} \frac{1}{(2i)} 
\left[e^{i(g+1) \varphi} -e^{i(g-1) \varphi} \right]  \, .
 \ee
For $g>1$, we integrate along $C + C_1$, where $C_1$ denotes the upper semicircle with radius $R_1 \to \infty$.  
Then, on the closed path we apply Cauchy's integral theorem. Since inside this path we have no singularity in the $\varphi$
plane, therefore the integral together with the limit $R_1 \to \infty$ is equal to zero.

For $g<1$ and since $g>0$ so that $(g+1)>1$, 
the first part of the integral (\ref{4.2}) vanishes and we are left with the clockwise path integral $C + C_2$:
\be
   \frac{-1}{\pi} \int\limits_{C + C_2} \frac{d \varphi}{\varphi} (\frac{1}{2i}) e^{-i(1-g) \varphi} \, .\nonumber			   
\ee                             
For $R_2 \to \infty$ we again obtain no contribution. Expanding the exponential in this expression we obtain
\be
   \frac{-1}{\pi}  \lk \frac{1}{2i} \right) \int\limits_{C + C_2} \frac{d \varphi}{\varphi}  \left[1-i(1-g) \varphi + \ldots \right] \, .  		   
             \nonumber                       \ee
Here we apply Cauchy's residue theorem, which gives
\be
   \frac{-1}{\pi} \lk \frac{1}{2i} \right) (-)2 \pi i \cdot 1 = 1 \hk \mbox{for} \hk  g<1 \, , \nonumber
\ee
and finishes our proof for (\ref{4.1}).
\bi

\no     
Furthermore, DIRICHLET uses EULER'S formula (which actually was later proved by Poisson)
\be
\label{4.3}
           \int\limits^\infty_0 dve^{iqv} v^{s-1}  = \frac{\Gamma(s)}{(\pm q)^s}  e^{\pm i(\frac{\pi}{2})s}    
                               \ee
with $0<s<1$ and where the upper (lower) sign holds for positive (negative) $q$ values. For points $(x,y,z)$
on the surface of  the ellipsoid with semiaxes $a,b,c$ we have
\be
                             g(x,y,z) = \frac{x^2}{a^2} + \frac{y^2}{b^2} + \frac{z^2}{c^2} = 1, \nonumber
                             \ee
and for points $(x,y,z)$ inside (outside) of the ellipsoid we have
$g(x,y,z)<1(>1)$.
The potential of the homogeneous ellipsoid at a point $(\xi,\eta,\zeta)$ 
is then given by the volume integral over the entire ellipsoid:
\be
\label{4.4}
V (\xi,\eta,\zeta) = G\rho \int\limits_V dxdydz \frac{1}{r} \, , \quad \quad  r^2= (x-\xi)^2 + (y-\eta)^2 + (z-\zeta)^2 \, ,
\ee
where $\rho$ is the constant density and $G$ denotes Newton's gravitational constant. 
Now with the aid of (\ref{4.1}) (not Dirichlet's variant!), we can extend the volume integral (\ref{4.4}) over the entire space:
	 \be
	 \label{4.5}
           V(\xi,\eta,\zeta) = \frac{G\rho}{\pi} \int\limits_C d\varphi \frac{(\sin \varphi)}{\varphi}  \int \int\limits^{\infty}_{- \infty} \int 
           \frac{dxdydz}{r} e^{i\varphi g(x,y,z)} \,    .
           \ee
For the representation of the inverse distance $r^{-1}$ we make use of Euler's formula (\ref{4.3}), with $s = \frac{1}{2}, q = r^2>0$
and $\Gamma(\frac{1}{2}) = \sqrt{\pi}$ so that
\be
                                 \frac{1}{r} = \frac{e^{-i \frac{\pi}{4}}}{\sqrt{\pi}}
                                 \int\limits^\infty_0 \frac{dv}{\sqrt{v}}  e^{ir^2 v} \, . \nonumber
\ee
The explicit expression for the potential is then given by
                      \be
                      \label{4.6}
 V(\xi,\eta,\zeta) = G\rho \frac{e^{\frac{-i\pi}{4}}}{\pi^{\frac{3}{2}}} \int\limits^{+ \infty}_0 
 \frac{dv}{\sqrt{v}} \int\limits_C d\varphi \frac{(\sin \varphi)}{\varphi} \int\limits^{+ \infty}_{- \infty} dxu(x,\xi) 
 \int\limits^{+ \infty}_{- \infty} 
 dyu(y,\eta) 
  \int\limits^{+ \infty}_{- \infty} 
 dzu(z,\zeta) \, ,   
                          \ee     	
where  
\be      
\int\limits^{+ \infty}_{- \infty} dxu(x,\xi) = \int\limits^{+ \infty}_{- \infty}  dx \exp[i(\frac{x^2}{a^2}) \varphi+i(x-\xi)^2 v]  
 = a\sqrt{\pi} \frac{e^{i \frac{\pi}{4}}}{\sqrt{\varphi+a^2 v}}  
 \exp [\frac{e^{iv\xi^2 \varphi}}{(\varphi+a^2v)}] \nonumber
\ee
with similar expressions for the integrals over $y$ and $z$.
After substituting these expressions into (\ref{4.6}) we obtain
\begin{align}
V(\xi,\eta,\zeta) & = iG\rho abc \int\limits^\infty_0 \frac{dv}{\sqrt{v}} \int\limits_C d \varphi \frac{(\sin\varphi)}{\varphi}  \times \nonumber\\
& \exp[iv\varphi (\frac{\xi^2}{(\varphi+a^2 v)} + \frac{\eta^2}{(\varphi+a^2 v)} + \frac{\zeta^2}{(\varphi+c^2 v)})] [(\varphi+a^2 v) 
( \varphi+b^2 v)( \varphi+c^2 v)]^{- \frac{1}{2}} \nonumber
\end{align}
Replacing the integration variable $v$ by $v = \frac{\varphi}{\lambda}$ with fixed $\varphi$, we finally obtain
\be
\label{4.7}
                V(\xi,\eta,\zeta) = iG\rho abc \int\limits^\infty_0 \frac{d \lambda}{\sqrt{\Psi (\lambda)}}
                \int\limits_C d\varphi \frac{(\sin\varphi)}{\varphi^2}
                e^{iS\varphi}
                \ee
                      with	 
                      \be
                      \label{4.8}
                      \Psi(\lambda) = [(a^2+ \lambda)(b^2+\lambda)(c^2+\lambda)]
\ee
 and 
 \be
 S(\xi,\eta,\zeta;\lambda) = \frac{\xi^2}{(a^2+\lambda)} + \frac{\eta^2}{(b^2+\lambda)} + \frac{\zeta^2}{(c^2+\lambda)} \, .\nonumber 
\ee
The integral over $C$ now replaces DIRICHLET's  integration  over the positive real semi-axis where the use of 
EULER's formula (\ref{4.3}) fails. If one performs the partial derivatives of (\ref{4.7}) 
with respect to $\xi, \eta,  \zeta$, only the first power of $\varphi$ remains in the denominator of the integral, and formula
(\ref{4.1}) becomes applicable (compare F. HOPFNER [6]), whereas with DIRICHLET's integration the implementation of (\ref{4.3}) 
is not permitted, according to the above-mentioned exclusion of the NEWTONian case of $p = 2$. 
According to DIRICHLET's method, the potential could only be treated for $p > 4$
and the attraction components only for $p > 2$. 
Since in the first use of EULER's formula in (\ref{4.3}) the limit $1 < p < 3$
has already been included, DIRICHLET has to forgo the derivation of the potential 
and limit himself to the attraction components for $2 < p < 3$. 
The complex formulation removes this difficulty and leads to  
\be
\label{4.9}
i \int\limits_C d\varphi \frac{(\sin \varphi)}{\varphi^2}  e^{iS\varphi}  = \frac{1}{2} \int\limits_C \frac{d\varphi}{\varphi^2} 
(e^{i(S+1)\varphi}  -e^{i(S-1)\varphi}) \, .
\ee
Similarly to the computation of (\ref{4.2}), 
we now have to discuss the cases $S>1$ and $S<1$. 
Again, we employ Cauchy's integral theorem and close the path $C_2$ 
clockwise in the lower complex half plane and so obtain for (\ref{4.9})
\begin{align}
\label{4.10}
\frac{1}{2} \int\limits_{C + C_2} \frac{d \varphi}{\varphi^2} e^{- i (1 - S) \varphi} &=
\frac{1}{2} \int\limits_{C + C_2} \frac{d \varphi}{\varphi^2} (1 - i (1 - S) \varphi - \ldots) \nonumber\\
                           & =\frac{1}{2} 2\pi i (-i)(1-S) = \pi(1-S). 
 \end{align}
 We conclude that if the test point $(\xi,\eta,\zeta)$ lies inside the ellipsoid, 
$S(\xi,\eta,\zeta;\lambda)<1$, we find for the potential
                             \be
                             \label{4.11}
V_i(\xi,\eta,\zeta;\lambda) =  G\rho abc \pi \int\limits^\infty_0 \frac{d\lambda}{\sqrt{\Psi(\lambda)}}
(1- S(\xi,\eta,\zeta;\lambda)) \, , \quad \quad    S(\xi,\eta,\zeta;\lambda)<1 \, .
                                         \ee
If the test point lies outside, there exists exactly one value $\lambda = u$
that denotes the positive real root of $S(\xi,\eta,\zeta;\lambda) = 1$; 
for all other values $\lambda>u, S(\xi,\eta,\zeta;\lambda)$ is smaller than $1$. 
The lower value $u$ is a function of $\xi,\eta,\zeta$
of the test point. Therefore the potential of the ellipsoid in external space is given by
\be
\label{4.12}
            V_e = G\rho abc \pi \int\limits^\infty_u \frac{d\lambda}{\sqrt{\Psi(\lambda)}}  (1- S(\xi,\eta,\zeta;\lambda)) 		      \, .
                                          \ee
Finally let us apply our formula to the simple case, namely the potential of a homogeneous sphere with radius $R = a = b = c$.
Admittedly, this case belongs to a first-semester course in mechanics. However, it illustrates very nicely Dirichlet's path 
to reproducing Newton's result.
We start with
\be
                      S(\xi,\eta,\zeta;\lambda) = (r^2+ \lambda)^{-1} (\xi^2 + \eta^2 + \zeta^2) < 1,
\nonumber
\ee
and the inner point is given by $r^2 = \xi^2 + \eta^2 + \zeta^2$.
So we are given  $S (\xi,\eta,\zeta;\lambda) = \frac{1}{(R^2+\lambda)} r^2$ 
and $\left[ (a^2 + \lambda)( b^2 + \lambda)(c^2 + \lambda) \right]^{\frac{1}{2}}  = (R^2 +\lambda)^{3/2}$.

According to formula (\ref{4.11}), the potential for a test point $r$ inside the massive homogeneous sphere is given by
\begin{align}
\label{4.13}
V_i (r)  & = G \rho R^3 \pi \int\limits^\infty_0 d \lambda \frac{\left[ 1 - \frac{r^2}{(R^2 +\lambda)} \right]}{(R^2 + \lambda)^{3/2}} = 
G\rho R^3 \pi \int\limits^\infty_0 d \lambda \frac{(R^2 +\lambda -r^2)}{(R^2 + \lambda)^{5/2}}
                   \nonumber\\
            &        =  G \rho R^3 \pi 
                               \int\limits^\infty_0 d \lambda 
                                \left[ \frac{1}{( R^2 +\lambda)^{3/2}} -
                                \frac{r^2}{(R^2 + \lambda)^{5/2}} \right] \, .
                           \end{align}
The two integrals can easily be calculated so that we get
\be
\label{4.14}
V_i (r)  = G \rho R^3 \pi \left[  \frac{2}{R} -
\frac{2}{3} \frac{r^2}{R^3} \right]= G \rho \pi \frac{2}{3}  \left[ 3 R^2 - r^2 \right] \,.                  (4.14)
\ee
For the potential of the external point we first have to determine the lower limit $u$, 
which follows from $S(\xi,\eta,\zeta;u) = 1 = \frac{r^2}{(R^2 +u)}$, i.e.,
$u = r^2 - R^2$. This requires the value of the integral, as in (\ref{4.12}):
\begin{align}
 V_e (r)  & = G \rho R^3 \pi \left[ \int\limits^\infty_{r^2 - R^2} d \lambda \frac{1}{(R^2+\lambda)^{3/2}}
 - r^2 \int\limits^\infty_{r^2 - R^2} d\lambda \frac{1}{(R^2 +\lambda)^{5/2}} \right] \,  
\nonumber\\
&  =  G \rho R^3 \pi \left[ \frac{2}{r} - \frac{2}{3} (\frac{1}{r} ) \right] = 
G \rho(\frac{4}{3} \pi R^3) \frac{1}{r} = G \rho V (\frac{1}{r}) = G \frac{M}{r} \, . \hk \mbox{(Newton)} \, .
\end{align}

\section{Acknowledgement}

I enjoyed many discussions with Nils Schopohl, with whom I shared my interest in Dirichlet's work.



\bibliographystyle{plain}
\def\cprime{$'$}

\end{document}